\documentclass[twocolumn]{jpsj3sc}

\usepackage{txfonts}
\usepackage{color}

\title{Mechanism of High-Temperature Superconductivity in Correlated-Electron Systems
}

\author{Takashi Yanagisawa}

\inst{
National Institute of Advanced Industrial Science and Technology
1-1-1 Umezono, Tsukuba, Ibaraki 305-8568, Japan
}

\abst{
It is very important to elucidate the mechanism of superconductivity for 
achieving room temperature superconductivity.
This paper is a short review article on the mechanism of high-temperature
superconductivity.
In the first half of this paper,
we give a brief review on mechanisms of superconductivity in many-electron systems.
We believe that high-temperature superconductivity
may occur in a system with interaction of large-energy scale.
Empirically, this is true for superconductors that have been found so far.
In the second half of this paper, we discuss cuprate high-temperature
superconductors.
We argue that superconductivity of high temperature cuprates is
induced by the strong on-site Coulomb interaction, that is, the origin
of high-temperature superconductivity is the strong electron correlation. 
We show the results on the ground state of electronic models for high temperature cuprates
on the basis of the optimization variational Monte Carlo method.
A high-temperature superconducting phase will exist in the 
strongly correlated region.}

\begin{document}
\maketitle

\section{Introduction}
It is a challenging research subject to clarify the mechanism of high 
temperature superconductivity, and~indeed it has been studied intensively
for more than 30 years~\cite{bed86,kei15,ryb16}. 
For this purpose, it is important to clarify the ground state and phase diagram
of electronic models with strong correlation because high temperature cuprates
are strongly correlated electron~systems.

Most   superconductors induced by the electron--phonon interaction have
$s$-wave pairing symmetry.  We can understand physical properties 
of $s$-wave superconductivity based on the Bardeen--Cooper--Schrieffer
(BCS) theory~\cite{coo56,bar57,bcs}.
The critical temperature $T_c$ of most of electron--phonon superconductors
is very low except for exceptional compounds.
Many unconventional superconductors  that cannot be
understood by the BCS theory  have been discovered.
They are, for~example, heavy fermion superconductors, organic
superconductors and cuprate superconductors for which the pairing
mechanism is different from the electron--phonon interaction.
In particular, cuprate superconductors exhibit relatively high $T_c$
and have become of great interest.
A common feature in both electron--phonon systems and correlated
electron systems is that critical temperature may have a strong
correlation with the energy scale of the interaction that induces
electron~pairing.

This paper has two parts.  In~the first part, we give a review on  mechanisms
of superconductivity in the electron--phonon system and in the
correlated electron system.
In the second part, we mainly discuss the mechanism of high-temperature
cuprates.

The model for  CuO$_2$ plane in cuprate superconductors is called the 
d-p model or the
three-band Hubbard
model~\cite{eme87,hir89,sca91,ung93,ogu94,koi00,yan01,koi01,yan03,koi03,koi06,yan09,
web09,lau11,web14,ave13,ebr16,tam16}.
It is certainly a very difficult task to elucidate the phase diagram of the d-p model.
Simplified models are also used to investigate the mechanism of
superconductivity, for~example  the two-dimensional (2D) single-band
Hubbard model~\cite{hub63,hub64,gut63,cep77,gro87,yok87,gia91,zha97,zha97b,yan96,
yan95b,nak97,yam98,yam00,yam11,
har09,yan13a,bul02,yok04,yok06,aim07,miy04,yan08,yan13,yok13,yan16}
and  ladder model~\mbox{\cite{noa95,noa97,yam94,yan95,nak07,koi99}}.
The Hubbard model was introduced to understand
the metal--insulator transition~\cite{mot74} and was employed to
understand various magnetic phenomena~\cite{mor85,yos96}.
On the basis of the Hubbard model, it is possible to understand the 
appearance of inhomogeneous states
reported for cuprates, such as 
stripes~\cite{tra96,suz98,yamd98,ara99,moc00,wak00,bia96,bia13} and
checkerboard-like density wave states~\cite{hof02,wis08,han04,miy09}.
It was also expected that the Hubbard model can account for high
temperature superconductivity~\cite{and87}.

A variational Monte Carlo method is used to examine the ground
state properties of strongly correlated electron systems, where we
calculate the expectation values exactly using a numerical
method~\cite{cep77,gro87,yok87,gia91,nak97,yam98,yam00,yam11,har09,yan13a}.~We introduced the wave function of $\exp(-S)$-type in the study of
superconductivity in the Hubbard model~\cite{yan98,yan14,yan19}.
This wave function is very excellent in the sense that the
energy expectation value is lower than that of any other
wave functions~\cite{yan16}.

The paper is organized as follows.  In~{Section \ref{sec2}, we discuss the phonon 
mechanism of superconductivity.  In~Section~\ref{sec3}, we discuss the electron
mechanism of superconductivity. Section~\ref{sec4} is devoted to a discussion
on superconductivity in correlated electron systems. 
In Section~\ref{sec5}, we show the model for high temperature cuprates.
We present the optimization variational Monte Carlo method (OVMC) in
Section~\ref{sec6}.  We show the results on superconductivity based on the OVMC in Section~\ref{sec7}.
We discuss the stability of antiferromagnetic state in Section~\ref{sec8}.
We show the phase diagram when the hole doping rate is changed in Section~\ref{sec9}.
We give a summary in Section~\ref{sec10}}.

\section {Part I. Superconductivity in Many-Electron Systems}

\subsection{Possibility of High-{$T_c$} Superconductivity}\label{sec2}

In the BCS theory, the~electron--phonon
interaction is assumed to induce attractive interaction between
electrons and the pairing symmetry is $s$-wave~\cite{coo56,bar57,bcs}. 
There are many superconductors with $s$-wave pairing symmetry and
most of them are due to the electron--phonon interaction.
The BCS theory was successful to explain physical properties of
these~superconductors.

In the strong-coupling theory based on the Green function
formulation~\cite{eli60,car90}, the~critical temperature $T_c$ was 
estimated as~\cite{mcm68},
\begin{equation}
T_c = \frac{\theta_D}{1.45}\exp\left( 
-\frac{1.04(1+\lambda)}{\lambda-\mu^*(1+0.62\lambda)}\right),
\end{equation}
where $\lambda$ is the electron--phonon coupling constant,  
$\theta_D=\hbar\omega_D/k_B$ is the Debye temperature and $\mu^*$
is the renormalized Coulomb parameter defined by
\begin{equation}
\mu^*  = \frac{\mu}{1+\mu\ln(\epsilon_F/\omega_D)},
\end{equation}
for $\mu=U/\epsilon_F$ where $U$ is the strength of the Coulomb
interaction and $\epsilon_F$ is the Fermi energy.
$\mu^*$ is the phenomenological parameter being approximately~0.1.
 
The electron--phonon coupling constant $\lambda$ is expressed as
\begin{equation}
\lambda = 2\int_0^{\infty}d\omega \frac{\alpha(\omega)^2F(\omega)}{\omega},
\end{equation}
where $\alpha(\omega)$ is the averaged electron--phonon coupling
over the Fermi surface and $F(\omega)$ indicates the product
of the spectral function of phonon and the density of states.
This is approximately written as
\begin{equation}
\lambda \simeq \frac{\rho(\epsilon_F)\langle I\rangle^2}{M\omega_D^2},
\end{equation}
where $\rho(\epsilon_F)$ is the density of states at the Fermi surface
and $M$ is the mass of an atom.
McMillan predicted that $T_c$ would have a limit being of the
order of 30~K from the analysis for this formula\cite{mcm68}. 

The McMillan formula was modified by replacing $\theta_D/1.45$ by 
logarithmic Debye frequency $\omega_{\ln}$ where~\cite{all75}
\begin{equation}
\omega_{\ln}= \exp\left( \frac{2}{\lambda}\int_0^{\infty}d\omega
\alpha(\omega)^2F(\omega)\frac{\ln \omega}{\omega}\right).
\end{equation}

It was predicted that high critical temperature would be
possible for large $\lambda$ since $T_c\propto \sqrt{\lambda}$
for $\lambda \gg 1$.
If $\omega_{\ln}$ is large, $\lambda$ is also large, and~the
crystal is stable, high $T_c$ would be realized.
It was predicted that high $T_c$ would be realized in 
hydrogen solid with high Debye temperature~\cite{ash68}. 
In fact, high temperature superconductors with $T_c$ above
200 K were discovered under extremely high pressure (160$\sim$200 GPa)
in hydrogen compounds such as H$_3$S and LaH$_{10}$ \cite{dro15,pen17,liu17}.

It is important to consider multi-band superconductors in the search
for high temperature superconductors.  In~fact, MgB$_2$ and iron
based superconductors are multi-band superconductors. 
An important role of Lifshitz transition in iron based superconductors
and MgB$_2$ multi-band superconductors has been predicted~\cite{inn10}.
An interesting point is that the possibility of high-$T_c$ superconductivity
in materials where tuning the chemical potential shows a quasi-1D
Fermi surface topology as in organics and hydrides~\cite{maz17}.
A layered superconductor such as cuprate superconductor can be 
regarded as a multiband
superconductor due to interlayer couplings.
A multi-band superconductivity has been investigated as a generalization
of the BCS theory since early works on the two-band 
superconductivity~\cite{mos59,suh59,per62,kon63}.
There will appear many interesting properties in
superconductors with multiple gaps such as time-reversal symmetry
breaking~\cite{sta10,tan10a,tan10b,dia11,yan12b,hu12,sta12,pla12,mai13,wil13,gan14,yer15,hil09,has09},
the existence of massless modes~\cite{yan13c,lin12,kob13,koy14,yan14b,tan15},
unusual isotope effect~\cite{val97,cho09,shi09,yan09b,per12}
and fractional-flux quantum
vortices~\cite{izy90,vol09,kup11,tan18,yan18}.
When we have multiple order parameters, there appear multiple
Nambu--Goldstone bosons and Higgs
bosons~\cite{yan13c,lit82,cea14,pek15,cea15,koy16,yan17b,yan17,ait99,mur17,yan18b}.
This will result in significant excitation modes that are unique
in multi-band~superconductors.

It is important to include in a theoretical picture the presence of
multiple electronic components with anomalous normal state properties
in the charge and spin sector, e.g.,~the well known Fermi arcs and
charge pseudogap phenomenology.
The  ``shape resonance''  scenario of multigap BCS-BEC crossover has been
proposed~\cite{per96,bia98}.
The study of the electronic structure of the cuprates superconductors
Bi$_2$Sr$_2$CaCuO$_{8+y}$ and La$_2$CuO$_{4+y}$ doped by mobile
oxygen interstitials using local probes has shown a scenario made of
two electronic components: a strongly correlated Fermi liquid which
coexists with stripes made of anisotropic polarons condensed into a 
generalized Wigner charge density wave~\cite{kus00,mul98,bia94}.

\subsection{Electron Correlation and~Superconductivity}\label{sec3}

We discuss the electron correlation due to the Coulomb interaction
between electrons.
The on-site Coulomb interaction is important in the study of the
metal insulator transition and magnetic properties of materials.
The Hubbard model is written as~\cite{hub63}
\begin{equation}
H= \sum_{ij\sigma}t_{ij}c^{\dag}_{i\sigma}c_{j\sigma}
+U\sum_in_{i\uparrow}n_{i\downarrow},
\end{equation}
where $t_{ij}$ indicates the transfer integral and the second term
denotes the Coulomb interaction with the strength  $U$.
$t_{ij}$ are chosen as follows.   $t_{ij}=-t$ when $i$ and $j$ are
nearest-neighbor pairs $\langle ij\rangle$
and $t_{ij}=-t'$ when $i$ and $j$ are next-nearest neighbor pairs.
In the following, $N$ is the number of lattices, and~$N_e$ denotes
the number of~electrons.

When two electrons {spin up and down} at the same site, 
the energy becomes higher by $U$ where $U$ denotes the on-site
Coulomb energy.
In the case of half-filling, the~Mott transition occurs when
$U(>0)$ is as large as the bandwidth and the ground state is 
an insulator. 
The effective Hamiltonian is derived in the limit of large 
$U/t$ \cite{har67,cha77,cha78},
based on the canonical transformation $H_{{\rm eff}}= e^{iS} He^{-iS}$.
In the limit $U/t\rightarrow\infty$, the~double occupancy
is not allowed.
The effective Hamiltonian is written as
\begin{equation}
H_{{\rm eff}}= H+i[S,H]+\frac{i^2}{2}[S, [S,H]]+\cdots.
\end{equation}

We write the Hamiltonian as $H=\tilde{H}_0+H_{1}$ where
\begin{eqnarray}
\tilde{H}_0&=& \sum_{ij\sigma}t_{ij}(a_{i\sigma}^{\dag}a_{j\sigma} 
+d_{i\sigma}^{\dag}d_{j\sigma})+U\sum_in_{i\uparrow}n_{i\downarrow},\\
H_{1}&=& \sum_{ij\sigma}t_{ij}(a_{i\sigma}^{\dag}d_{j\sigma}
+d_{j\sigma}^{\dag}a_{i\sigma}).
\end{eqnarray} 

Here, we defined $a_{i\sigma}=c_{i\sigma}(1-n_{i,-\sigma})$ and
$d_{i\sigma}=c_{i\sigma}n_{i,-\sigma}$.
$a_{i\sigma}=c_{i\sigma}(1-n_{i.-\sigma})$ is the electron operator
without double occupancy.
We choose $S$ to satisfy $i[S,\tilde{H}_0]+H_{1}=0$, so that $H_{{\rm eff}}$
reads in the subspace of no double occupancy,
\begin{equation}
H_{{\rm eff}}= \sum_{ij\sigma}t_{ij}a_{i\sigma}^{\dag}a_{j\sigma}
+\frac{i}{2}[S,H_1]+\frac{i^2}{3}[S,[S,H_1]]
+\cdots.
\end{equation}

When we consider only the nearest-neighbor transfer $t_{ij}=-t$,
the effective Hamiltonian reads
\begin{eqnarray}
H_{{\rm eff}}&=& -t\sum_{\langle ij\rangle\sigma}a_{i\sigma}^{\dag}a_{j\sigma}
-\frac{t^2}{U}\sum_{j\mu\mu'}\big[ 
a_{j+\mu\uparrow}^{\dag}a_{j\downarrow}^{\dag}a_{j\downarrow}a_{j+\mu'\uparrow}
\nonumber\\ 
&+& a_{j\uparrow}^{\dag}a_{j+\mu\downarrow}^{\dag}a_{j+\mu'\downarrow}a_{j\uparrow}
+a_{j+\mu\uparrow}^{\dag}a_{j\downarrow}^{\dag}a_{j+\mu'\downarrow}a_{j\uparrow}
+a_{j\uparrow}^{\dag}a_{j+\mu\downarrow}^{\dag}a_{j\downarrow}a_{j+\mu\uparrow}\big],
\nonumber\\
\end{eqnarray}
where $j+\mu$ and $j+\mu'$ denote the
nearest-neighbor sites in the $\mu$ and $\mu'$ directions, respectively.
The second term being proportional to $t^2/U$ contains the 
nearest-neighbor exchange interaction
and also three-site terms
when $\mu\neq \mu'$.  The~three-site terms are of the same order as the
exchange interaction.  When we neglect the three-site terms, the~effective Hamiltonian reduces to the t-J model given by
\begin{equation}
H_{{\rm eff}} = -\sum_{\langle ij\rangle\sigma}(a_{i\sigma}^{\dag}a_{j\sigma}
+{\rm h.c.})+  J\sum_{\langle ij\rangle}\left({\bf S}_i\cdot {\bf S}_j
-\frac{1}{4}\tilde{n}_i\tilde{n}_j\right),
\end{equation}
where $J=4t^2/U$ and $\tilde{n}_i=\tilde{n}_{i\uparrow}+\tilde{n}_{i\downarrow}$ 
with $\tilde{n}_{i\sigma}=a_{i\sigma}^{\dag}a_{i\sigma}$.

High-temperature cuprates and heavy fermion systems are typical
correlated electron systems and many superconductors have been 
reported.
Most of superconductors in these systems have nodes in the
superconducting gap, namely,  the~Cooper pair is anisotropic.
This indicates that superconductivity is unconventional and 
does not conform to the conventional BCS theory. 
The mechanism of superconductivity is certainly non-phonon
mechanism.  
We show several characteristic properties of cuprate high-temperature
superconductors:
\begin{enumerate}
\item[1.]  The~Cooper pair has $d$-wave~symmetry.
\item[2.]  The~superconducting phase exists near the antiferromagnetic
phase and parent materials are a Mott~insulator.
\item[3.]  The~CuO$_2$ plane is commonly contained and the on-site Coulomb
repulsive interaction works between $d$ electrons.
\item[4.]  The~size of Cooper pair is very small being of order of 2\AA.
\item[5.]  The~CuO$_2$ plane is high anisotropic and there is a weak Josephson
coupling between two~layers.
\end{enumerate}

The small size of Cooper pair also supports the non-phonon
mechanism of cuprate superconductivity~\cite{mic90,rob93,arr91,kap12}.
A plausible non-phonon mechanism is due to the
Coulomb interaction on the same atom.
Because the energy scale of the Coulomb interaction is very large,
which is of the order of eV, we can expect superconductivity
with high critical temperature $T_c$. 
The critical temperature of heavy fermion materials is, however,
very low,  although~superconductivity occurs due to strong
Coulomb interaction between $f$ electrons.
This is because the effective mass of $f$ electrons is very large
in heavy fermion systems owing to the large self-energy correction.
The effective mass enhancement of heavy fermion materials becomes
as large as 100--1000, which means that the effective cutoff
becomes very small.  As~a result, the~characteristic energy scale is 
reduced considerably and the critical temperature $T_c$ becomes
very low begin of the order of 1 K.
In heavy fermion systems, the~characteristic energy scale is
given by the Kondo temperature $T_K$.
The ratio of the effective mass $m^*$ to the band mass $m_0$ is
approximately given as $m^*/m_0\simeq D/T_K$ for the bandwidth
$D$ and $T_K$.
Thus, the~effective bandwidth for heavy fermions is given by
the Kondo temperature $T_K\simeq D/(m^*/m_0)$.
Empirically, $T_c$ is lowered as the effective mass increases.
This is expressed as follows:
\begin{equation}
k_BT_c \simeq 0.1t/(m^*/m_0),
\end{equation}
where $t$ denotes the transfer integral proportional to the
bandwidth.
The estimated values of the transfer $t$, the~ratio $m^*/m_0$ and $T_c$
for several compounds are shown in {Table \ref{tab1}.} 
For cuprates, the~transfer $t$ is estimated as $t\sim 0.51$ eV.
The bandwidth for iron pnictides is about  five  times smaller than that for cuprates.
A list of typical superconductors in correlated electron systems
is shown in {Table \ref{tab2}.}

\begin{table}
\caption{The transfer integral $t$, effective mass $m^*$ and critical temperature
$T_c$ in correlated electron systems, where  $m_0$ denotes the band mass.
The orders of these quantities are shown in the table.
For heavy fermions, $t/(m^*/m_0)$ corresponds to the Kondo temperature $T_K$.
For Hydrides, the~Debye frequency $\omega_{\ln}$ is shown.  }
\centering
\begin{tabular}{cccccc}
\hline
\textbf{ }  & \boldmath{$t$ or $\omega_{\ln}$}  & \boldmath{$m^*/m_0$} & \boldmath{$t/(m^*/m_0)$} & \boldmath{$T_c$} &  \\
\hline
Cuprates        &  5000 K      &  5         &  1000  &  100 K  & 
$t\sim 0.51$eV~\cite{fei96} \\
Fe pnictides     &  1000 K      &  $\sim 2$  &  500   &   50 K & $t\sim 0.1$eV~\cite{kur08}\\
Heavy fer    &  10000 K     &  100$\sim$1000  &  10$\sim$100  & 1$\sim$10 K &
 \cite{ste91,hew93,onu18} \\
Organics         & 200$\sim$500 K  &  2$\sim$5  &  100  &  10 K &  \cite{ish12} \\
Hydrides         &  1000 K      & $\sim 1$  & 1000  & 100 K & $\omega_{\ln}$ \cite{aka15} \\
\hline
\end{tabular}
\label{tab1}
\end{table}

\begin{table}
\caption{Superconducting materials.
}
\centering
\begin{tabular}{ccccc}
\hline
\textbf{Materials}  & \boldmath{$T_c$}  & \textbf{Pair Symmetry} & \textbf{Refs} \\
\hline
CeCu$_2$Si$_2$ & 0.6 K  &   $s$ or $d$     &   \cite{ste79,sak14} \\
UPt$_3$             & 0.52 K &   $p$ or $f$ &    \cite{ste84}\\
UBe$_{13}$          & 0.86 K  &  $p$     &     \cite{ott83} \\
URu$_2$Si$_2$ & 1.2 K &               &     \cite{pal85,ami99,ohk99} \\  
CeRu$_2$           & 6.2 K   &   $s$      &    \cite{hed97}\\
UPd$_2$Al$_3$  &  2 K    &  $d$      &   \cite{gei91,kyo93,ina99} \\
UNi$_2$Al$_3$  &  1 K    &  $p$~?     &   \cite{kyo93,ish02}\\
CeCoIn$_5$        &  2.3 K  &  $d$  &    \cite{pet01,iza01} \\
CeRhIn$_5$        &  2.1 K  &  $d$  &   \cite{heg00} \\
                       & (16.3 kbar) &       &           &  \\
CeRh$_2$Si$_2$ & 0.35 K &      &       \cite{mov96}  \\
                         &  (9 kbar)                  &        &     \\
UGe$_2$          & 0.8 K   &  $p$~?   &    \cite{sax00} \\
                      & (13.5 kbar)  &                    \\
URhGe            & 0.25 K  &  $p$~?    &   \cite{aok01}\\
Sr$_2$RuO$_4$  & 1.5 K  &   $p$ or $f$ &   \cite{mae94}\\
PrOs$_4$Sb$_{12}$ & 1.85 K &    line nodes~?  &  \cite{bau02}\\
Na$_{{\rm x}}$CoO$_{2-{\rm y}}$$\cdot$H$_2$O & 5 K &  $p$~? &   \cite{tak03} \\
Ba$_{1-{\rm x}}$K$_{\rm x}$BiO$_3$ &  30 K  &    $s$  &   \cite{cha85} \\
MgB$_2$            & 39 K   &  $s$   &  \cite{nag01}\\
La$_{2-{\rm x}}$Sr$_{\rm x}$CuO$_4$ & 36 K &  $d$ & \\
YBa$_2$Cu$_3$O$_{6+{\rm x}}$  & 90 K &  $d$ &  \\
Tl$_2$Ba$_2$Ca$_{{\rm n}-1}$Cu$_{\rm n}$O$_{2{\rm n}+4}$ & 125 K &  $d$ & \\
HgBa$_2$Ca$_{{\rm n}-1}$Cu$_{\rm n}$O$_{2{\rm n}+2+\delta}$ & 135 K &  $d$ &  \\
LaO$_{1-{\rm x}}$F$_{\rm x}$FeAs  &  26 K  &    &    \cite{kam08}  \\
NdFeAsO$_{1-{\rm y}}$  &  54 K  &    &    \cite{kit08}  \\
H$_3$S   &  203 K  &  s  &    \cite{dro15}  \\
LaH$_{10}$  & 260 K &  s  &   \cite{pen17,liu17,dro18,som18} \\
\hline
\end{tabular}
\label{tab2}
\end{table}

\subsection{Superconductivity in Strongly Correlated Electron~Systems}\label{sec4}

The possibility of superconductivity in strongly correlated electron
systems has been discussed intensively.  
The perturbative calculations such as the fluctuation-exchange
approximation (FLEX) have been performed to investigate the
superconducting ground state~\cite{bic89,pao94,mon94}.
There were, however, the~results by
quantum Monte Carlo methods, which did not support the existence of
high-temperature superconductivity in the two-dimensional Hubbard
model~\cite{zha97,zha97b,aim07}.
In quantum Monte Carlo calculations, the~strength of the Coulomb
interaction $U$ is not large enough because the range of accessible $U$
is very restricted.
It is now certain that there is a superconducting phase in the
strongly correlated region~\cite{yan16}.
The simplest wave function of superconducting state with strong
electron correlation is the Gutzwiller-projected BCS wave function:
\begin{equation}
\psi_{BCS-G} = P_G\prod_{\bf k}(u_{\bf k}+v_{\bf k}
c_{{\bf k}\uparrow}^{\dag}c_{-{\bf k}\downarrow}^{\dag})|0\rangle,
\end{equation}
where $u_{\bf k}$ and $v_{\bf k}$ are BCS parameters and $P_G$ is
the Gutzwiller operator to control the on-site electron
correlation.  $P_G$ is written as
\begin{equation}
P_G= \prod_{j}(1-(1-g)n_{j\uparrow}n_{j\downarrow}),
\end{equation}
where $g$ is a variational parameter in the range of
$0\le g\le 1$.
The ratio of $u_{\bf k}$ and $v_{\bf k}$ is given as
\begin{equation}
\frac{v_{\bf k}}{u_{\bf k}}= \frac{\Delta_{\bf k}}
{\xi_{\bf k}+(\xi_{\bf k}^2+\Delta_{\bf k}^2)^{1/2}},
\end{equation}
where $\xi_{\bf k}$ denotes the electron dispersion relation
measured from the Fermi energy and $\Delta_{\bf k}$ is the gap
function.  We use the following form for the gap function in the
two-dimensional case:
\begin{eqnarray}
\Delta_{\bf k} &=& \Delta( \cos k_x-\cos k_y)~~~ d-{\rm wave},\\
\Delta_{\bf k} &=& \Delta( \cos k_x+\cos k_y)~~~ {\rm anisotropic}-s-{\rm wave},\\
\Delta_{\bf k} &=& \Delta ~~~~~~~~~~ {\rm isotropic}~s-{\rm wave}.
\end{eqnarray}
$\Delta$ is a constant and is treated as a variational parameter.
The wave function $\psi_{BCS-G}$ is just the wave function that
Anderson proposed as a wave function of the resonate-valence-bond 
(RVB) state~\cite{and87}.

It has been shown that the ground-state energy has a minimum
at finite $\Delta$ for the BCS-Gutzwiller wave function with
$d$-wave symmetry in the two-dimensional Hubbard model by
using the variational Monte Carlo method~\cite{yam98}. 
The superconducting condensation energy $E_{cond}$ per site
in the limit of large system size was estimated as~\cite{yam98,yam00}
\begin{equation}
E_{cond}/N\simeq 0.2~{\rm  meV},
\end{equation}
where the transfer integral $t$ is set at 0.5 eV.
The similar result was obtained for the three-band d-p model~\cite{yan09}.
Thus, the~condensation energy per atom is of the order of $10^{-4}$ eV.

The superconducting condensation energy $E_{cond}$ for cuprate high-temperature
superconductors was evaluated by using the result of  specific heat 
measurement for YBCO as  0.17--0.26 meV per Cu atom~\cite{yam98,lor93}.
The estimation of $E_{cond}$ from the data of critical magnetic field
gives the similar result~\cite{hao91}.
The obtained results by theoretical calculations and experimental
measurements are very close each other. 
This agreement is very remarkable.
Thus, this value indicates
the characteristic energy  for cuprate high-temperature superconductors.
This result may support that the superconductivity in cuprate high
temperature superconductors is caused by the strong electron correlation
and the 2D Hubbard model includes essential~ingredients.\\

\section {Part II. Mechanism of Superconductivity in Cuprates}

We discuss the mechanism of superconductivity in this part.
We show numerical results obtained by using the optimized wave
functions.

\subsection{Model for High-\boldmath{$T_c$} Cuprates}\label{sec5}

The Hamiltonian of the d-p model for high-$T_c$ cuprates is
\begin{eqnarray}
H_{dp}&=& \epsilon_d\sum_{i\sigma}d_{i\sigma}^{\dag}d_{i\sigma}
+ \epsilon_p\sum_{i\sigma}(p_{i+\hat{x}/2\sigma}^{\dag}p_{i+\hat{x}/2\sigma}
+ p_{i+\hat{y}/2\sigma}^{\dag}p_{i+\hat{y}/2\sigma})
\nonumber\\
&+& t_{dp}\sum_{i\sigma}[d_{i\sigma}^{\dag}(p_{i+\hat{x}/2\sigma}
+p_{i+\hat{y}/2\sigma}-p_{i-\hat{x}/2\sigma}-p_{i-\hat{y}/2\sigma})
+ {\rm h.c.}]\nonumber\\
&+& t_{pp}\sum_{i\sigma}[p_{i+\hat{y}/2\sigma}^{\dag}p_{i+\hat{x}/2\sigma}
-p_{i+\hat{y}/2\sigma}^{\dag}p_{i-\hat{x}/2\sigma}
 - p_{i-\hat{y}/2\sigma}^{\dag}p_{i+\hat{x}/2\sigma}\nonumber\\
&& +p_{i-\hat{y}/2\sigma}^{\dag}p_{i-\hat{x}/2\sigma}+{\rm h.c.}]\nonumber\\
&+& t_d'\sum_{\langle\langle ij\rangle\rangle\sigma}\epsilon_{ij}
(d_{i\sigma}^{\dag}d_{j\sigma}
+{\rm h.c.} )
+ U_d\sum_i d_{i\uparrow}^{\dag}d_{i\uparrow}d_{i\downarrow}^{\dag}
d_{i\downarrow} \nonumber\\
&+& U_p\sum_i(n^p_{i+\hat{x}/2\uparrow}n^p_{i+\hat{x}/2\downarrow}
+n^p_{i+\hat{y}/2\uparrow}n^p_{i+\hat{y}/2\downarrow}).
\end{eqnarray} 

Since we use the hole picture in this paper,
$d_{i\sigma}$ and $d^{\dag}_{i\sigma}$ represent the operators for the $d$ hole.
$p_{i\pm\hat{x}/2\sigma}$ and $p^{\dag}_{i\pm\hat{x}/2\sigma}$ denote the
operators for the $p$ holes at the site $R_{i\pm\hat{x}/2}$, and~in a
similar way $p_{i\pm\hat{y}/2\sigma}$ and $p^{\dag}_{i\pm\hat{y}/2\sigma}$
are defined.
$n^p_{i+\hat{x}/2\sigma}$ and $n^p_{i+\hat{y}/2\sigma}$ are the
number operators of $p$ holes at $R_{i+\hat{x}/2}$ and $R_{i+\hat{y}/2}$,
respectively.
$t_{dp}$ is the transfer integral between adjacent Cu and O orbitals
and $t_{pp}$ is that between nearest p orbitals.
$t_d'$ indicates that between d orbitals where
$\langle\langle ij\rangle\rangle$ denotes a next nearest-neighbor pair of copper
sites.
$\epsilon_{ij}$ takes the values $\pm 1$ (see Figure 1).
This value is determined from the sign of the transfer
integral between next nearest-neighbor $d$ orbitals.
$U_d$ indicates the strength of the on-site Coulomb repulsion between
$d$ holes and $U_p$ is that between $p$ holes.

The values of band parameters were evaluated by several
works~\cite{web08,hyb89,esk89,mcm90,esk91}. 
We show an example: $U_d=10.5$, $U_p=4.0$
and $U_{dp}=1.2$ in eV~\cite{hyb89}.  Here,   $U_{dp}$ is the nearest-neighbor
 Coulomb interaction
between holes on adjacent Cu and O orbitals and is small compared to $U_d$.
$U_{dp}$ is neglected in this paper.
We write $\Delta_{dp}=\epsilon_p-\epsilon_d$.
The number of sites is denoted as $N$,
and the energy is measured in units of $t_{dp}$.

\vspace{2cm}
\begin{figure}
\begin{minipage}{0.5\hsize}
\includegraphics[width=6.0cm]{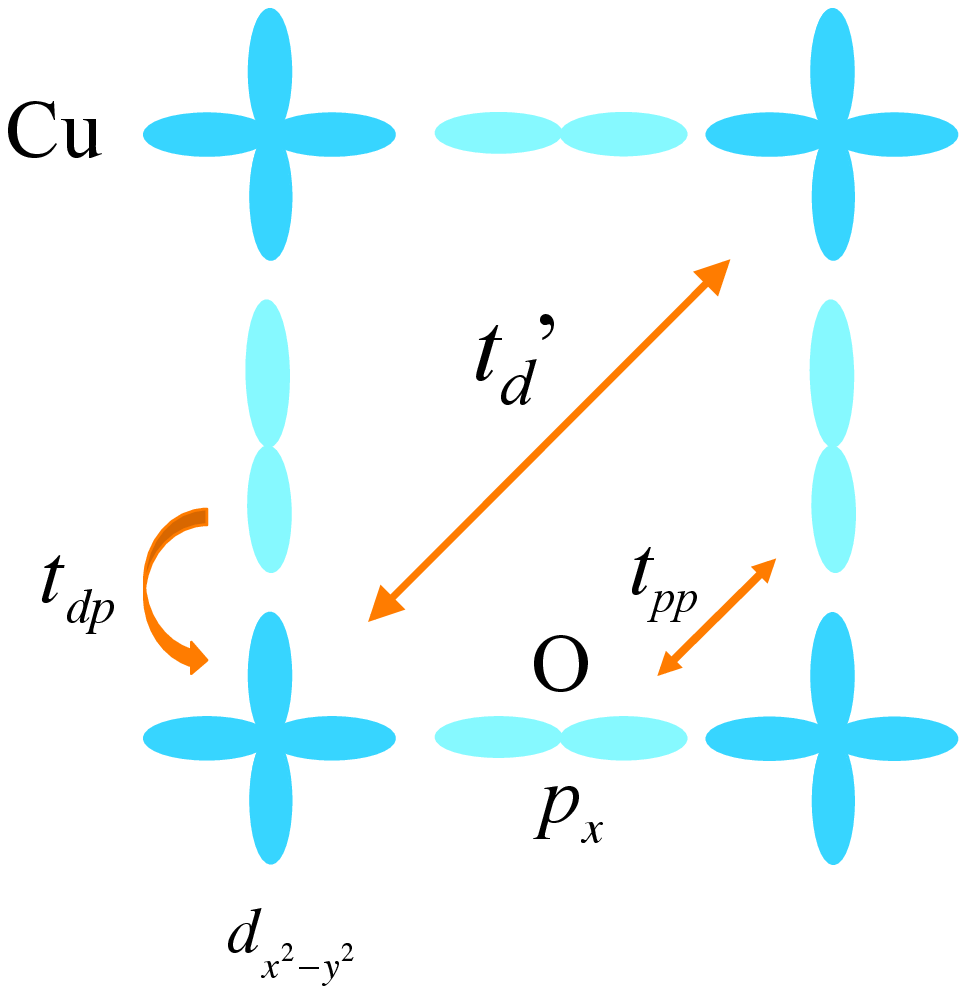}
\end{minipage}
\begin{minipage}{0.5\hsize}
\includegraphics[width=6.0cm]{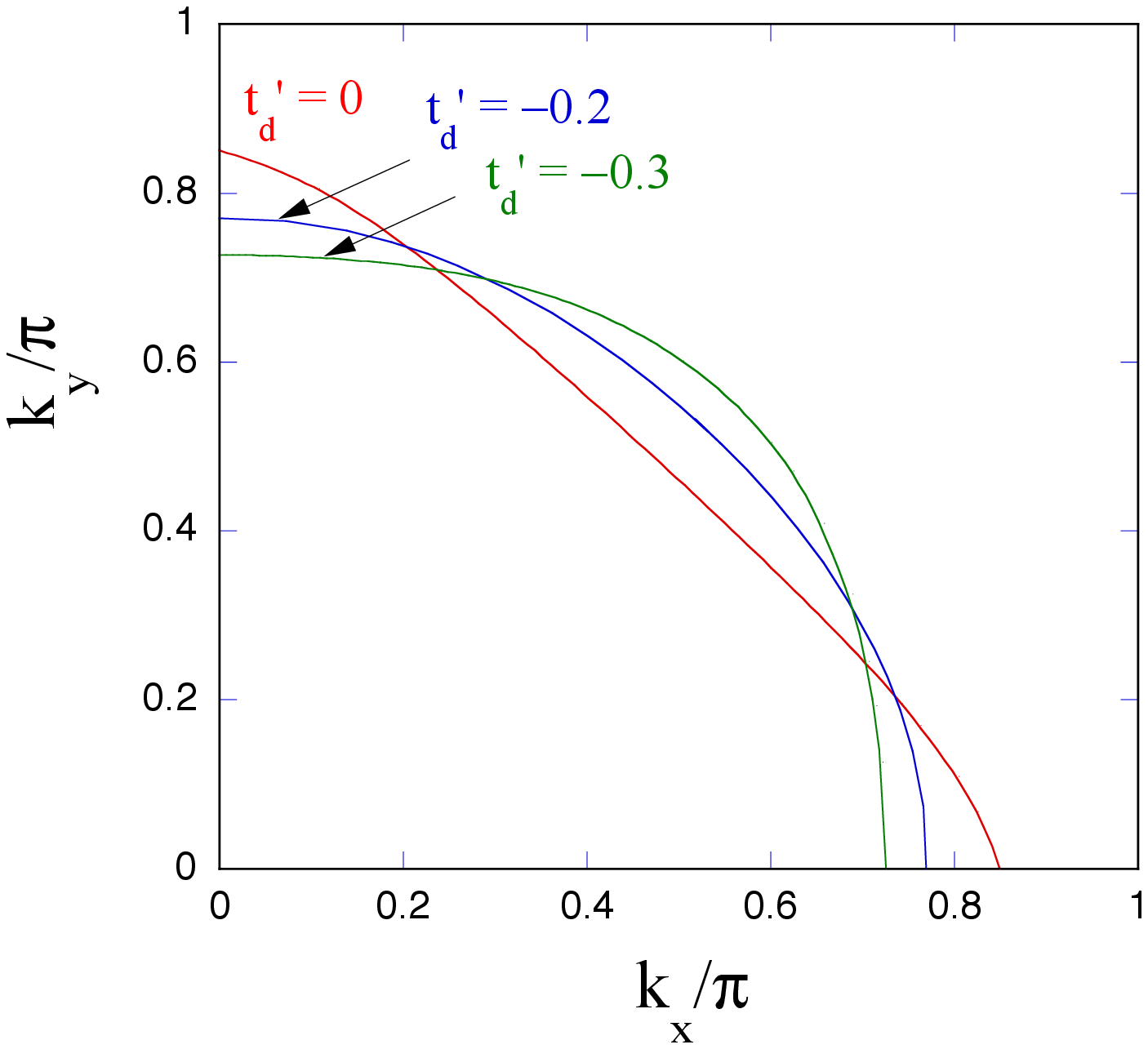}
\end{minipage}
\caption{
(\textbf{Upper}) The transfer integral $t_d'$ in the CuO$_2$ plane.
$t_{dp}$ and $t_{pp}$ are conventionally defined.
(\textbf{Lower})  Fermi surface of the d-p model
with the hole density 0.13~\cite{yan14}.
~We put $t_{pp}=0.4t_{dp}$ and
$\epsilon_p-\epsilon_d=2t_{dp}$ for $t_d'=0$, $-0.2t_{dp}$ and
$-0.3t_{dp}$.  
}
\end{figure}

\subsection{Optimization Variational Monte Carlo~Method}\label{sec6}

\subsubsection{Off-Diagonal Wave~Function}

The Gutzwiller wave function is
\begin{equation}
\psi_G = P_G\psi_0,
\end{equation}
where $\psi_0$
is a one-particle state.
Our purpose is to improve the Gutzwiller function.
We multiply the Gutzwiller function by an 
exponential-type operator.  The~wave function is given as
as~\cite{yan16,ots92,yan98,yan99,eic07,bae09,bae11}
\begin{equation}
\psi_{\lambda}= \exp(-\lambda K)\psi_G,
\label{wf1}
\end{equation}
where $K$ denotes the kinetic part of the Hamiltonian.  $\lambda$ is
a newly introduced real variational parameter~\cite{yan13a,yan98,yan99,yan07}.
There are other methods to improve the Gutzwiller function~\cite{yok04,mis14}.
The following Jastrow  operator is used~\cite{yok04},
\begin{equation}
P_{Jdh} = \prod_j\left( 1-(1-\eta)\prod_{\tau}\Big[
d_j(1-e_{j+\tau})+e_j(1-d_{j+\tau})\Big] \right),
\end{equation}
where $d_j$ is the operator for the doubly-occupied site given as
$d_j=n_{j\uparrow}n_{j\downarrow}$ and $e_j$ is that for
the empty site given by $e_j=(1-n_{j\uparrow})(1-n_{j\downarrow})$.
$\eta$ is the variational parameter in the range of
$0\le \eta\le 1$.  The~wave function is
\begin{equation}
\psi_{\eta}=P_{Jdh}\psi_G.
\end{equation}

In this paper,  we use the wave function of exponential type in
Equation~(\ref{wf1}) because 
the energy is further lowered when we use this wave function~\cite{yan16}.
The wave function for the d-p model is formulated similarly.
An initial state $\psi_0$ contains many variational parameters
($\tilde{t}_{dp}$, $\tilde{t}_{pp}$, $\tilde{t}'_d$, and~$\tilde{\epsilon}_p-\tilde{\epsilon}_d$):
\begin{equation}
\psi_0= \psi_0(\tilde{t}_{dp}, \tilde{t}_{pp}, \tilde{t}'_d,
\tilde{\epsilon}_p-\tilde{\epsilon}_d).
\end{equation}

We use $\tilde{t}_{dp}=t_{dp}$ as the energy unit.
We consider the following wave function that is improved from the
Gutzwiller wave function~\cite{yan16,yan98,ots92,yan99,eic07,bae09,bae11}:
\begin{equation}
\psi_{\lambda}= \exp(-\lambda K)\psi_G.
\end{equation}

The expectation values are evaluated by using the auxiliary
field method~\cite{yan98,yan07}.
The kinetic part $K$ also contains the band parameters $t_{pp}$,
$t'_d$ and $\epsilon_p-\epsilon_d$ as variational parameters:
\begin{equation}
K= K(\hat{t}_{pp}, \hat{t}'_d,
\hat{\epsilon}_p-\hat{\epsilon}_d).
\end{equation}

We take $\hat{t}_{pp}=\tilde{t}_{pp}$, $\hat{t}'_d=\tilde{t}'_d$ and
$\hat{\epsilon}_p-\hat{\epsilon}_d=\tilde{\epsilon}_p-\tilde{\epsilon}_d$,
for simplicity.
Thus,    we have $g$, $\tilde{t}_{pp}$, $\tilde{t}'_d$,
$\hat{\epsilon}_p-\hat{\epsilon}_d=\tilde{\epsilon}_p-\tilde{\epsilon}_d$,
and $\lambda$ as variational parameters.
The expectation values for this type of wave function are calculated
on the basis of the variational Monte Carlo method.  One can evaluate the
expectation value correctly within statistical~errors.

\subsubsection{Antiferromagnetic Wave~Function}

 The AF one-particle state $\psi_{AF}$ is formulated by the eigenfunction
of the AF trial Hamiltonian:
\begin{equation}
H_{AF}= \sum_{ij\sigma}t_{ij}c^{\dag}_{i\sigma}c_{j\sigma}
-\Delta_{AF}\sum_{i\sigma}(-1)^{x_i+y_i}\sigma n_{i\sigma},
\end{equation}
where $\Delta_{AF}$ is the AF order parameter and $(x_i,y_i)$
represents the coordinates of the site $i$.
With $\psi_{AF}$, the~wave function is given as
\begin{equation}
\psi_{\lambda,AF}= \exp(-\lambda K)P_G\psi_{AF}.
\end{equation}

\subsubsection{Superconducting Wave~Function}

We start from the BCS wave function
\begin{equation}
\psi_{BCS}= \prod_k(u_k+v_kc^{\dag}_{k\uparrow}c^{\dag}_{-k\downarrow})
|0\rangle,
\end{equation}
with coefficients $u_k$ and $v_k$ satisfying $u_k^2+|v_k|^2=1$.
We choose $u_k/v_k=\Delta_k/(\xi_k+\sqrt{\xi_k^2+\Delta_k^2})$ for
the gap function $\Delta_k$ and 
$\xi_k=\epsilon_k-\mu$.
We assume
$\Delta_k= \Delta_{SC}(\cos k_x-\cos k_y)$.
The Gutzwiller-projected BCS wave function is
\begin{equation}
\psi_{G-BCS}=P_{N_e}P_G\psi_{BCS},
\end{equation}
where $P_{N_e}$ indicates the operator to extract the state with $N_e$
electrons. 
The exponential-BCS wave function is given by
\begin{equation}
\psi_{\lambda}= e^{-\lambda K}P_G\psi_{BCS}.
\end{equation}

In this wave function, we perform the electron--hole transformation 
for down-spin electrons:
\begin{equation}
d_k= c^{\dag}_{-k\downarrow},~~~ d^{\dag}_k= c_{-k\downarrow};
\end{equation}
and not for up-spin electrons: $c_k= c_{k\uparrow}$.
The electron pair operator $c^{\dag}_{k\uparrow}c^{\dag}_{-k\downarrow}$
denotes the hybridization operator $c^{\dag}_kd_k$ in this~formulation.

\subsection{Correlated~Superconductivity}\label{sec7}

\begin{figure}
\begin{minipage}{0.5\hsize}
\includegraphics[width=7.0cm]{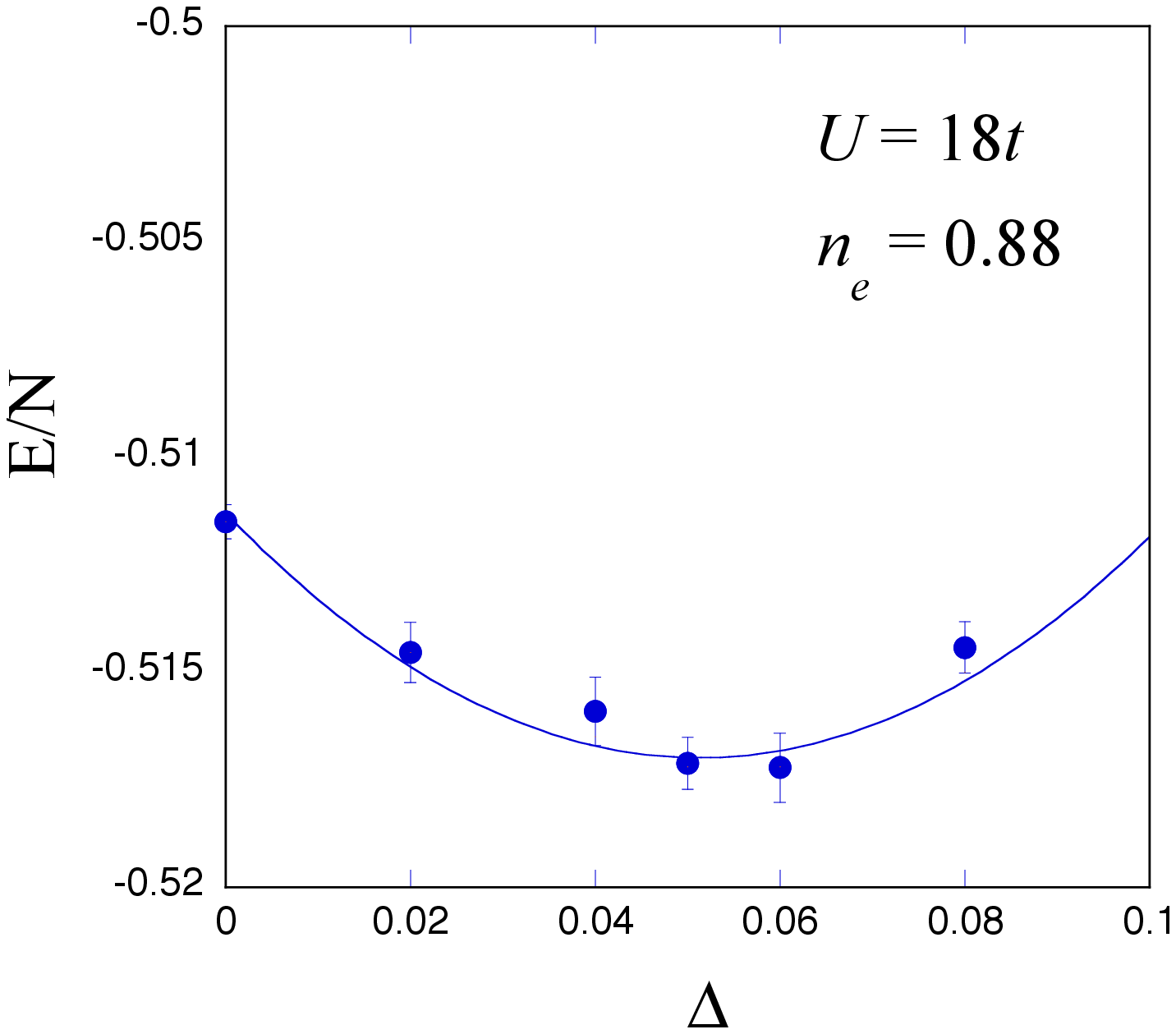}
\end{minipage}
\begin{minipage}{0.5\hsize}
\includegraphics[width=7.0cm]{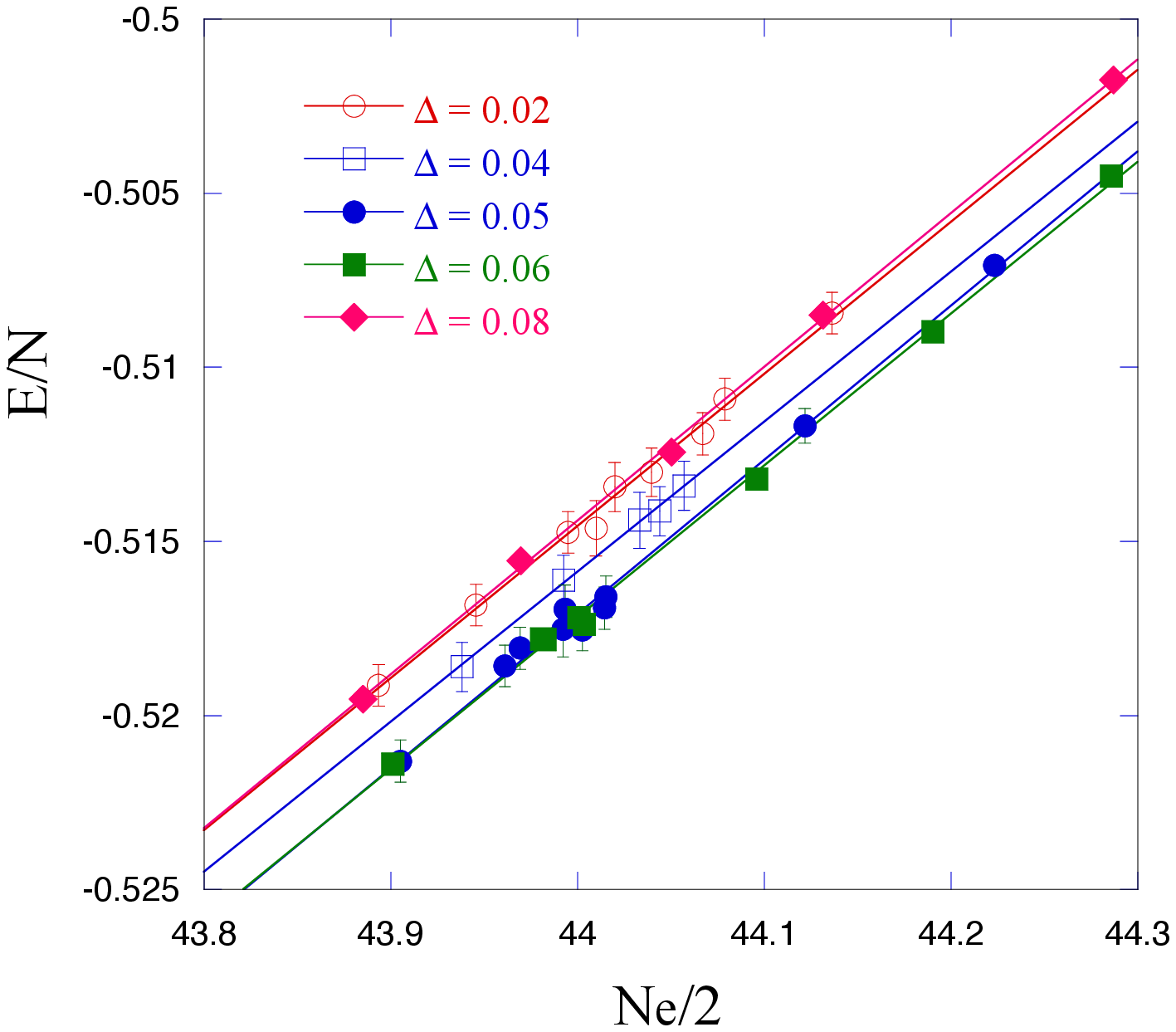}
\end{minipage}
\caption{
(\textbf{Upper}) The ground-state energy as a function of the superconducting
gap $\Delta$ for the optimized wave function $\psi_{\lambda}$ for the
Hubbard model on a $10\times 10$ lattice with $U/t=18$ and $t'=0$.
The electron density is $n_e=0.88$.
(\textbf{Lower})  The ground-state energy as a function of the electron
number where $\Delta$ is fixed for each line~\cite{yan16}.
}
\label{figure2}
\end{figure}

We first discuss the superconducting (SC) state in the two-dimensional
Hubbard model.  In~the optimization Monte Carlo method,  the~SC
state becomes indeed stable when the Coulomb interaction $U$ is
large to be of the order of the bandwidth. 
We show the ground-state energy as a function of the superconducting
order parameter $\Delta$ in {Figure \ref{figure2} (left).} 
The simple Gutzwiller-projected BCS wave function predicted the
possibility of superconductivity in the Hubbard model, and~the
improved wave function also shows a stability of the SC~state.

\begin{figure}
\centering
\includegraphics[width=6.5cm]{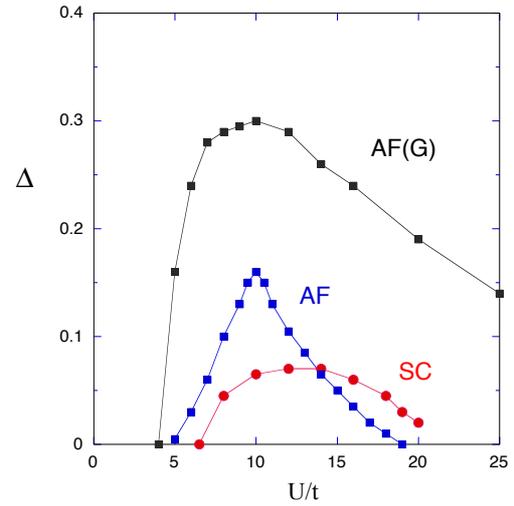}
\caption{
AF and SC order parameters as a function of
$U/t$ when $N_e=88$ for the 2D Hubbard model on a $10\times 10$ lattice.  
The periodic
boundary conditions are periodic in one direction and antiperiodic in the
other direction~\cite{yan16}.
AF(G) indicates the result obtained for the simple Gutzwiller
function.
}
\label{dE-sc-af}
\end{figure}

We show the SC and antiferromagnetic (AF) order parameters
as a function of $U$ in {Figure \ref{dE-sc-af}.}
The AF order parameter has a peak when $U/t\sim 10$,  which is
of the order of the bandwidth, and~the SC one also has a peak
at $U_c$ that is greater than the bandwidth. 
This indicates that there is the possibility of high-temperature
superconductivity in the strongly correlated~region.

\begin{figure}
\begin{minipage}{0.5\hsize}
\includegraphics[width=6.8cm]{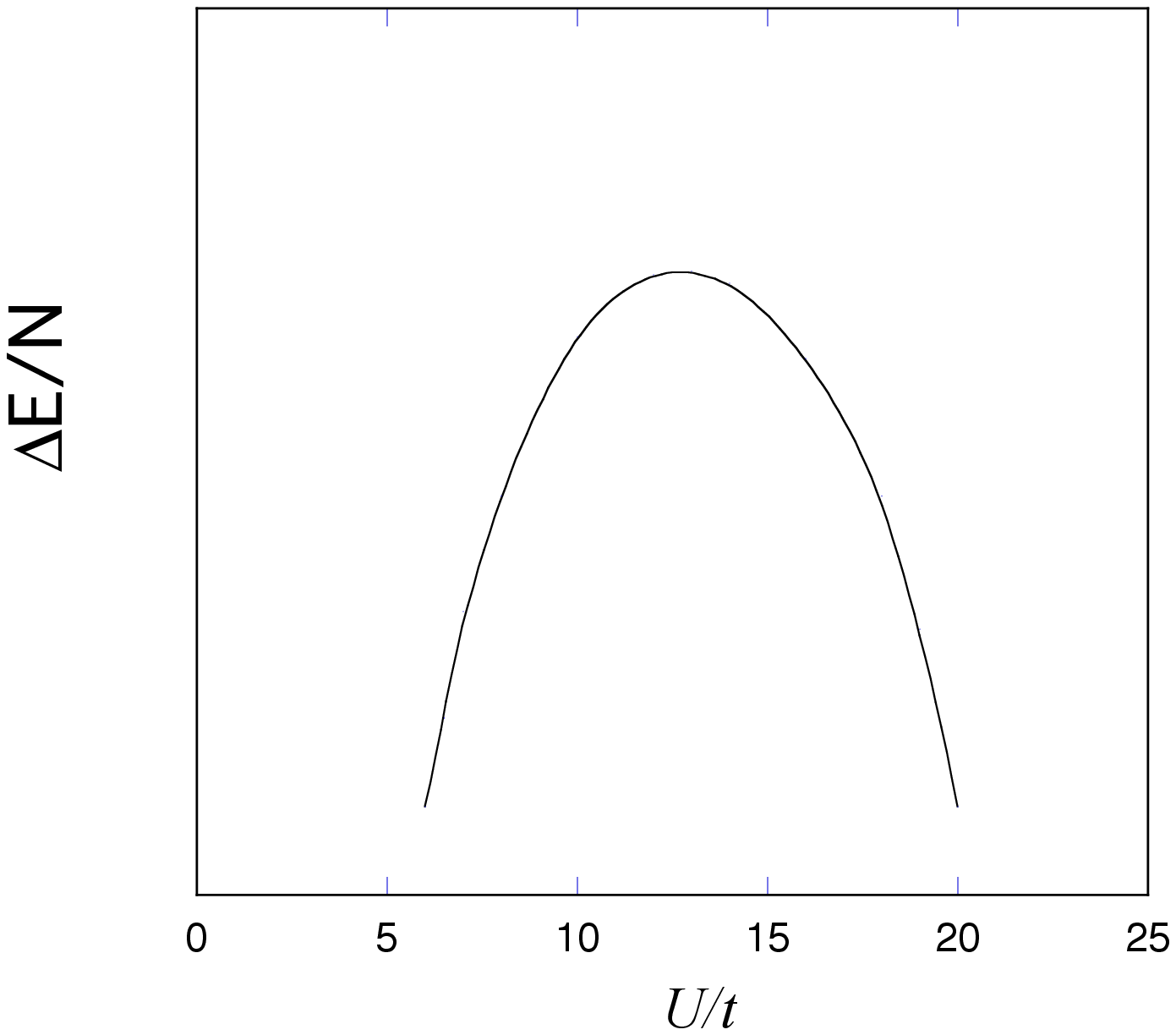}
\end{minipage}
\begin{minipage}{0.5\hsize}
\includegraphics[width=6.6cm]{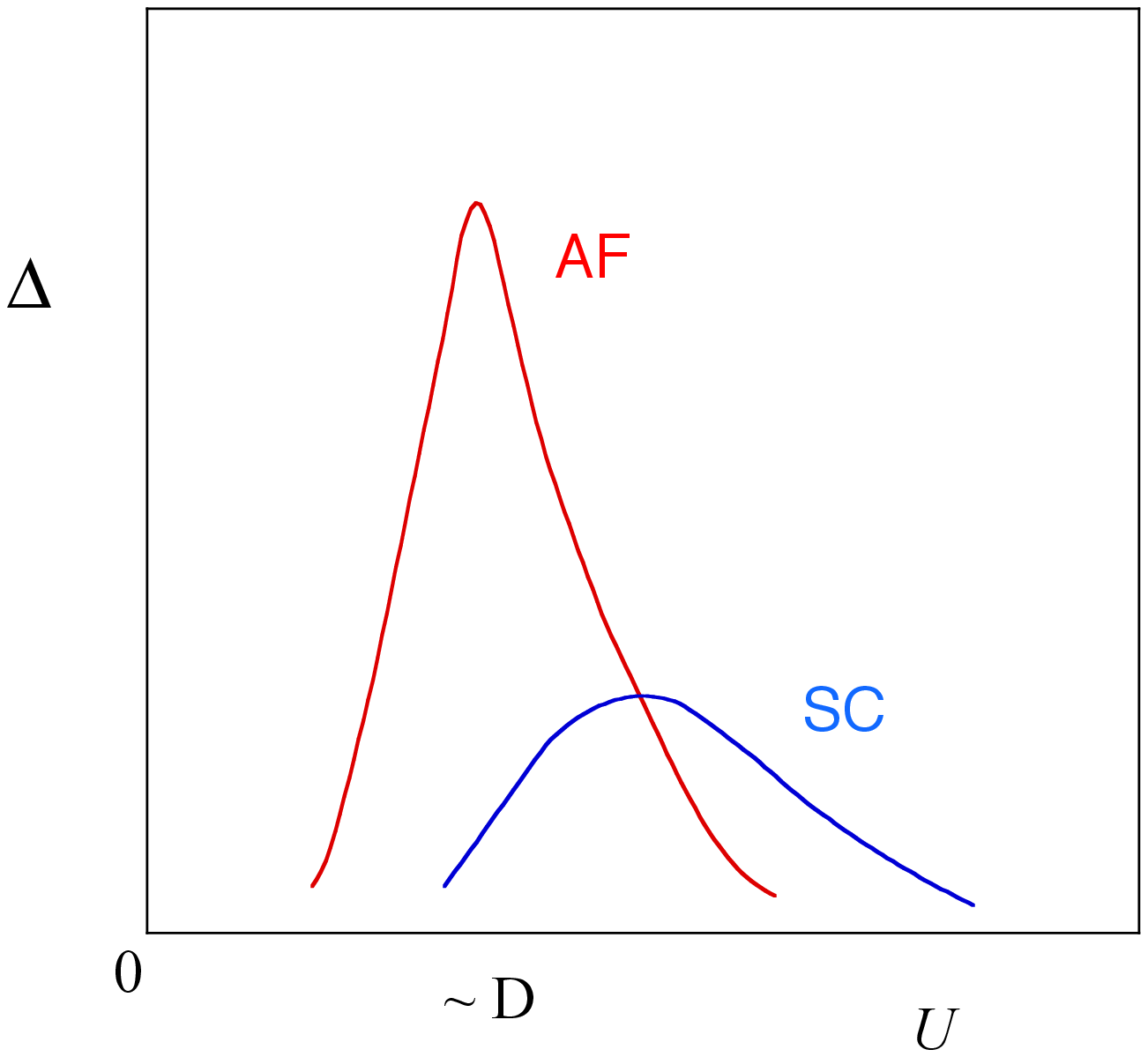}
\end{minipage}
\caption{
(\textbf{Upper})  A schematic picture of the superconducting condensation
energy as function of $U$ for the 2D Hubbard model.
(\textbf{Lower})  A schematic picture of the gap function of AF and AC states
as a function of $U$ for the 2D Hubbard model.
}
\label{figure4}
\end{figure}

The AF correlation is maximized at $U\sim U_c$ and decreases
when $U$ is larger than $U_c$.
We show schematic pictures in Figure~\ref{figure4}, where the SC condensation
energy as a function of $U$ is shown in the left panel, and~the AF and SC gap functions are shown in the right panel.
There is a crossover from weakly correlated region to the
strongly correlated region.  The~superconducting state is
most favorable when the AF correlation is gradually
suppressed in the strongly correlated region.
Thus, high temperature superconductivity is highly promising
in the strongly correlated region where $U$ is as large as
the bandwidth $D$ or larger than $D$.

\subsection{Stability of Antiferromagnetic~State}\label{sec8}

\subsubsection{Hubbard~Model}

\begin{figure}
\begin{minipage}{0.5\hsize}
\includegraphics[width=6.8cm]{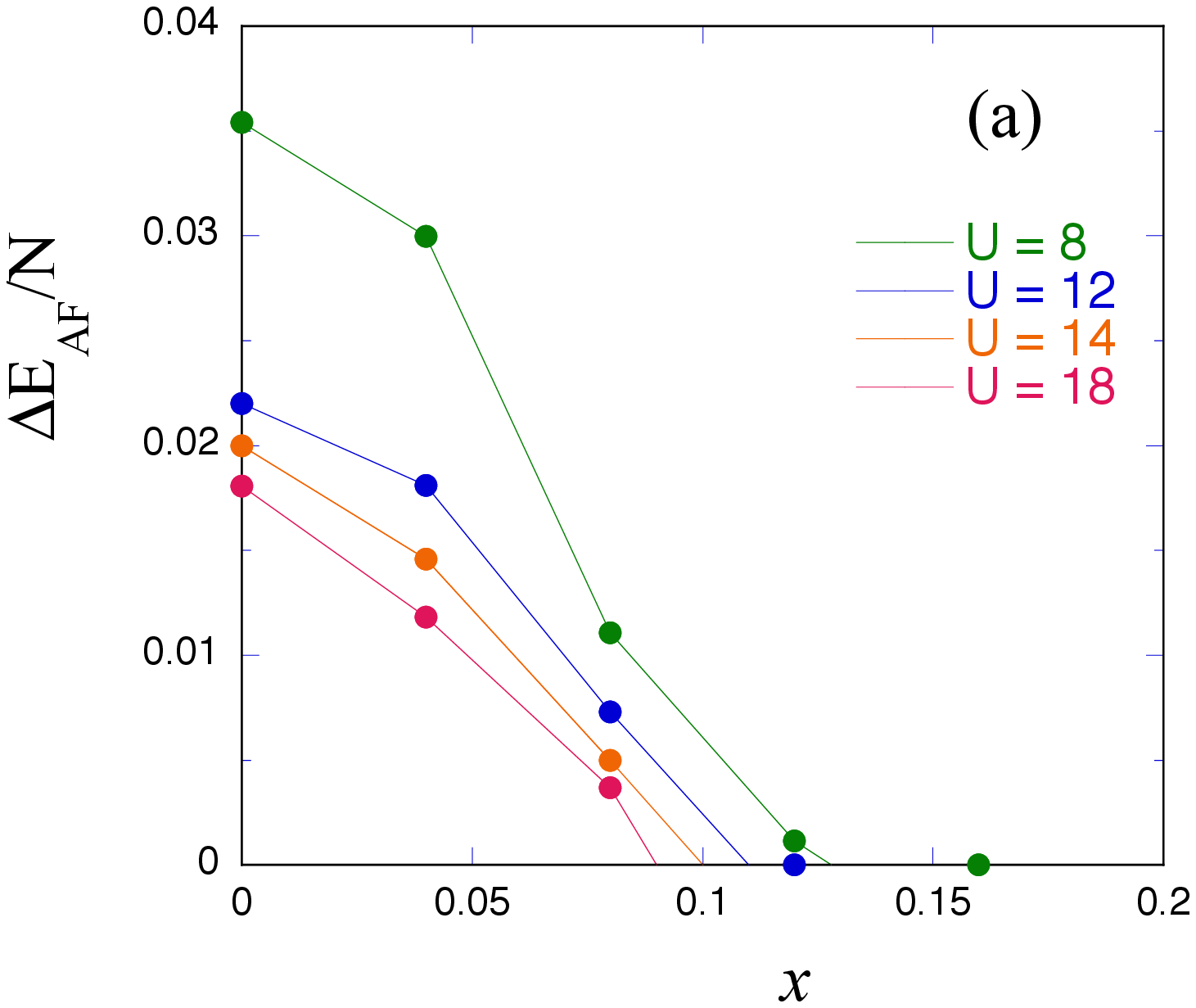}
\end{minipage}
\begin{minipage}{0.5\hsize}
\includegraphics[width=6.8cm]{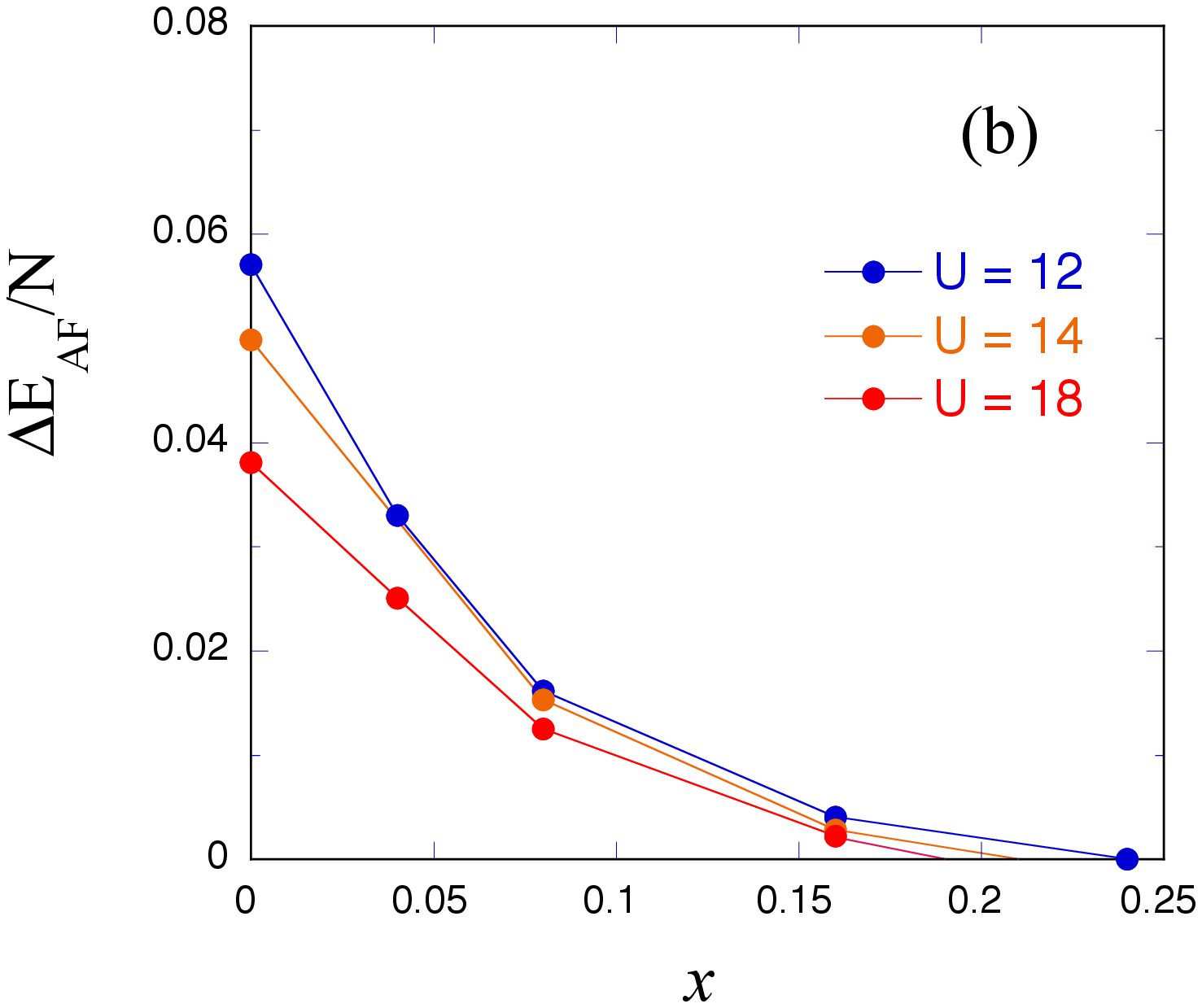}
\end{minipage}
\caption{
AF condensation energy $\Delta E_{AF}$ as a function of the hole doping rate
$x=1-n_e$ on a $10\times 10$ lattice for: $t'=0$ (\textbf{a}); and $t'=-0.2t$ (\textbf{b}) \cite{yan19}.
We put $U/t=12, 14$ and 18.
}
\label{fig5}
\end{figure}

Let us examine the stability of AF state.
There are two parameters $U$ and $t'$, and~there is the AF region
in the parameter space.  High temperature superconductivity is
expected near the boundary between the AF phase and the
paramagnetic phase.
We show the AF condensation energy $\Delta E_{AF}$ as a function of 
$1-n_e$ in Figure~\ref{fig5}a for $t'=0$
and Figure~\ref{fig5}b for $t'=-0.2t$.
The AF region becomes larger as $|t'|$ increases.
When $t'=-0.2t$, the AF region extends up to about 20$\%$ doping.
From the competition between superconductivity and AF order,
$t'=0$ is most favorable for superconductivity.

\subsubsection{Three-Band d-p~Model}

\begin{figure}
\centering
\includegraphics[width=6.6cm]{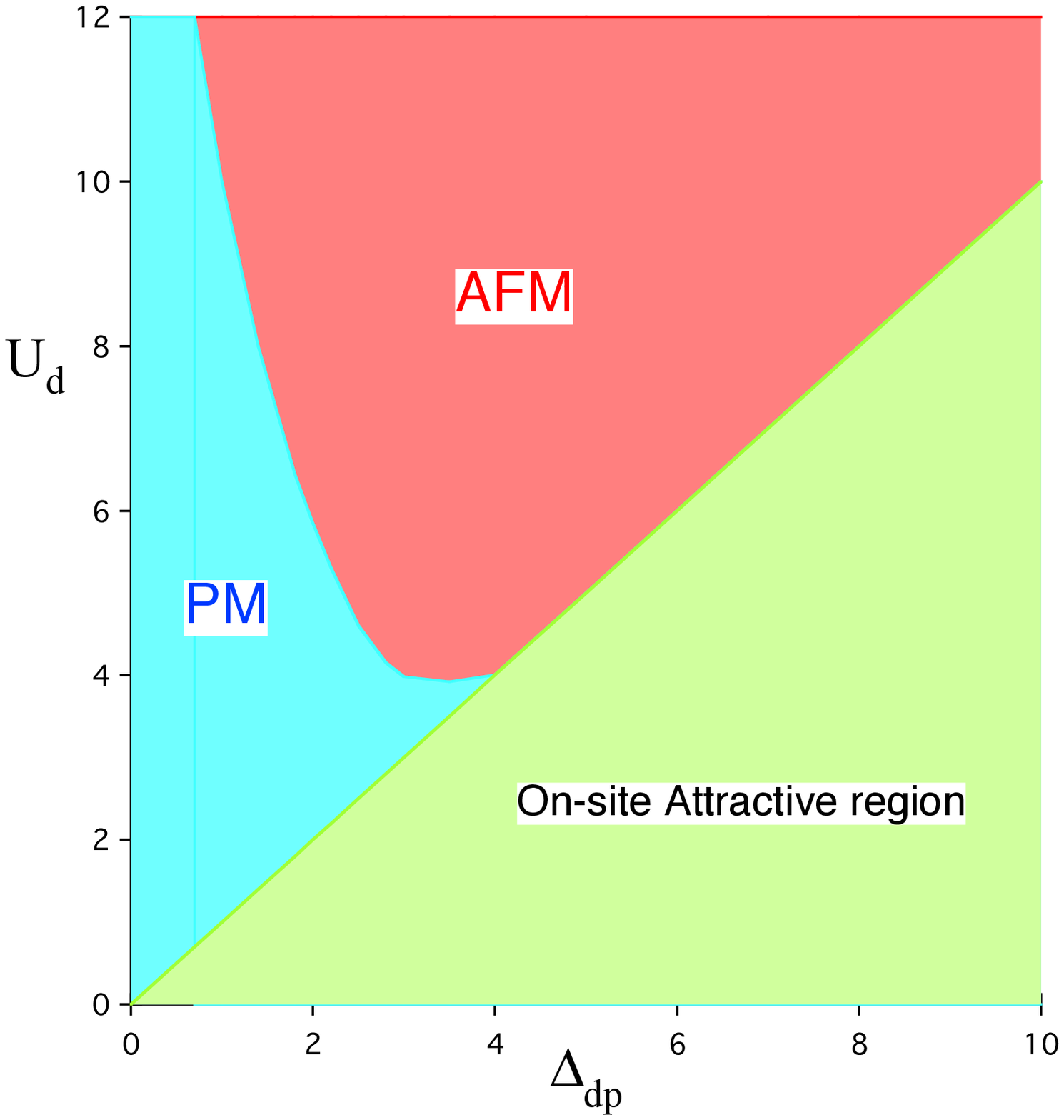}
\caption{
Antiferromagnetic and paramagnetic regions in the plane
of $U_{d}$ and $\Delta_{dp}=\epsilon_p-\epsilon_d$ for the d-p model.
We put $t_{pp}=0.4$ and $t_d'=0$.  There are 76 holes
on a $8\times 8$ lattice with 192 atoms in total.
The energy unit is given by $t_{dp}$. 
AFM and PM denote the antiferromagnetic metal and paramagnetic
metal, respectively.  There is a ``negative-U'' region when the
level difference is large where two $d$ electrons prefer to
occupy the same site.
The ground state may be a charge-density wave state or an $s$-wave 
superconducting state.  This is not clear yet.
}
\label{fig6}
\end{figure}

In general, in~the three-band d-p model, the~AF correlation is very strong
and the AF state is more stable than in the single-band Hubbard model. 
This is because $d$ electrons are localized and easily form magnetic
order~\cite{yan01}.
 To investigate the possibility of high temperature
superconductivity in the d-p model,
it is necessary to reveal regions with weak AF order.
There are many parameters in the d-p model to control the strength
of the AF correlation.  Among~them, the~Coulomb repulsion between
$d$ electrons $U_d$, the~level difference $\Delta_{dp}=\epsilon_p-\epsilon_d$,
and the hole density $x$ are important.  The~AF region is shown in Figure~\ref{fig6}
where $U_d$ and $\epsilon_p-\epsilon_d$ are varied, and~the hole
density is fixed at 0.1875.
The AF region increases when the hole density decreases.
We expect that
high temperature superconductivity will occur near the boundary
between AFM and PM regions.
This boundary exists in the region when $\Delta_{dp}$ is small.
High temperature superconductivity is likely occur when $\Delta_{dp}$
is small.
There is a  ``on-site attractive region'' when $\Delta_{dp}$ is large
where two $d$ electrons prefer to occupy the same site.
In this region,  a~charge-density wave or an $s$-wave superconducting
state will be~realized.

\begin{figure}
\centering
\includegraphics[width=6.6cm]{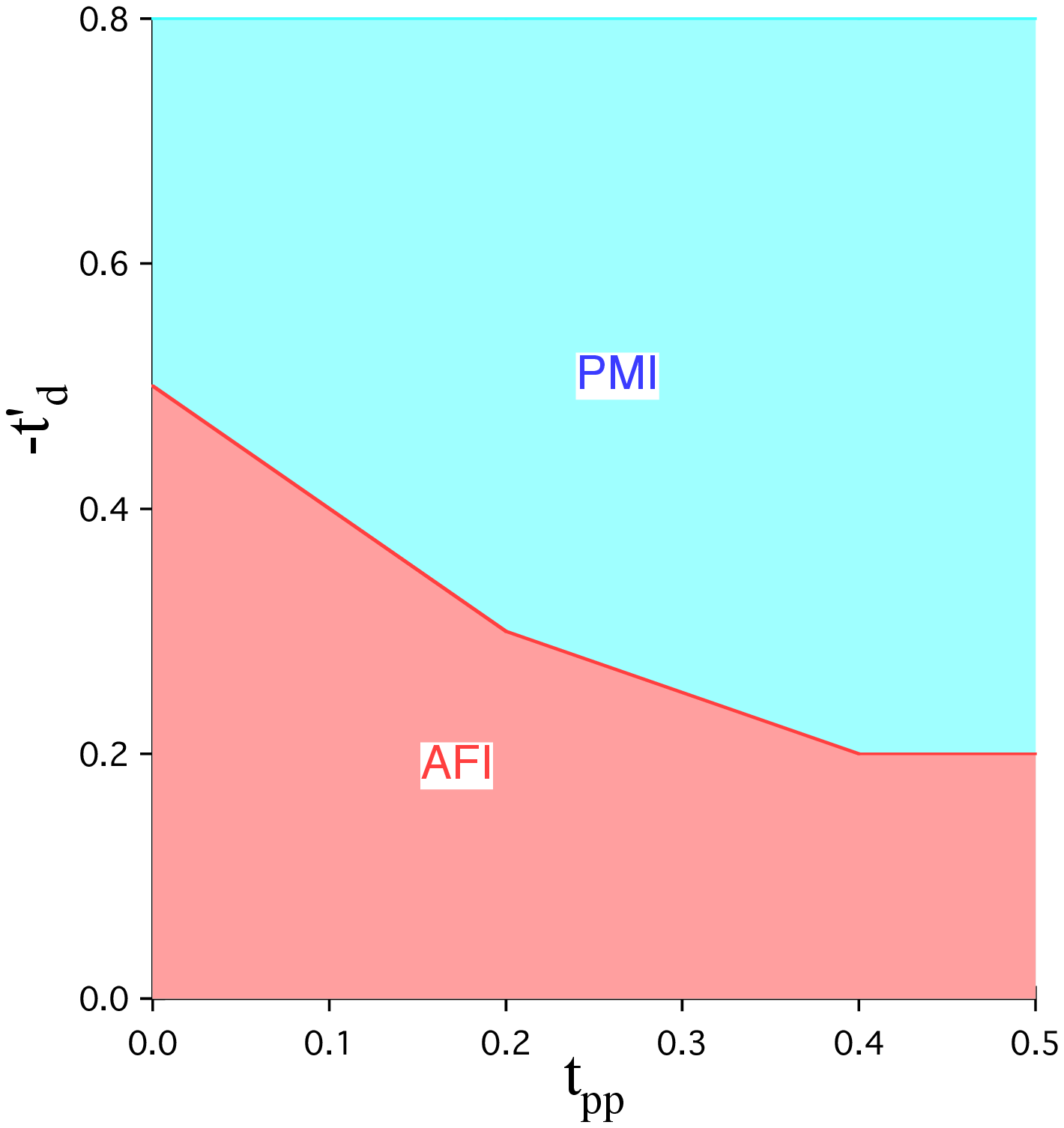}
\caption{
AF and paramagnetic insulator phases for the d-p model on a $6\times 6$ 
lattice~\cite{yan18c}.
Parameters are $U_d=8t_{dp}$, $U_p=0$, $\epsilon_p-\epsilon_d=t_{dp}$.
}
\label{fig7}
\end{figure}

We proposed to introduce the transfer
integral $t_d'$ to control the strength of AF correlation~\cite{yan14}.      
We show the AF region at half-filling in the $t_{pp}-t_d'$ plane in Figure~\ref{fig7}.
As $-t_d'$ increases, there is a phase transition from the AF insulator
to the paramagnetic insulator (PMI).
We expect that $t_d'$ and $t_{pp}$ will play an important role to
suppress AF correlation when holes are doped in the d-p~model.

\subsection{Phase Diagram for the Hubbard~Model}\label{sec9}

\begin{figure}
\centering
\includegraphics[width=8.0cm]{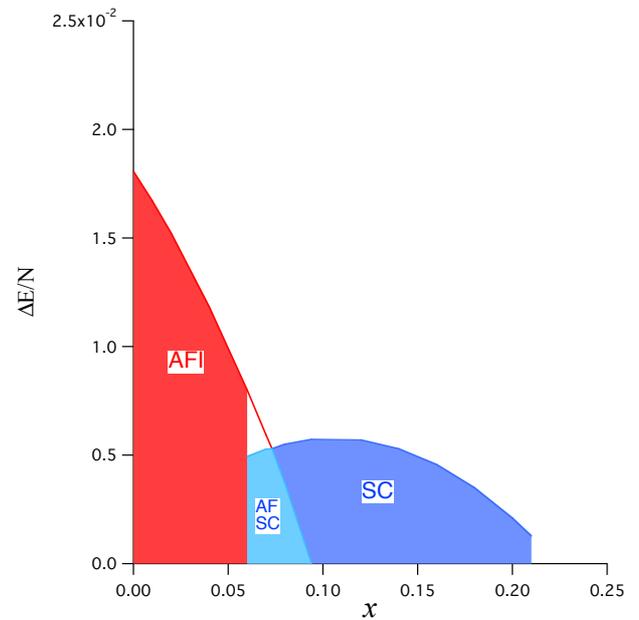}
\caption{
The condensation energy per site for the two-dimensional Hubbard model 
as a function of the hole doping rate $x$.
The calculation was carried out on a $10\times 10$ lattice.
AFI indicates the antiferromagnetic insulating state and SC denotes
the $d$-wave SC phase.  There is the coexistent state indicated as
AF-SC between these states.
Parameters are $t'=0$ and $U/t=18$.
}
\label{fig8}
\end{figure}

We discuss the phase diagram when carrier holes are doped
in the CuO$_2$ plane.  We evaluate the energy lowering when
we include the order parameter $\Delta$.  We define
\begin{equation}
\Delta E= E(\Delta=0)-E(\Delta_{min}),
\end{equation}
where $E(\Delta)$ takes a minimum at $\Delta=\Delta_{min}$.
We show $\Delta E$ as a function of the hole doping rate $x$ in Figure~\ref{fig8}
where we put $U/t=18$ and $t'=0$.
This phase diagram contains several interesting features.
There are three phases: antiferromagnetic insulator (AFI),
coexistent state (AFSC) and superconducting phase (SC).
When the hole doping rate $x$ is large,  e.g.,~$x>0.09$, the~pure
$d$-wave stat is stable.  
There is the possibility of high (and room) temperature superconductivity
in this phase.
In the underdoped region, approximately
$0.06<x<x_{dSC}$ with $0.08<x_{dSC}<0.09$, there is the coexistent state of
antiferromagnetism and superconductivity.  This is the mixed phase
of AF and SC.
$x_{dSC}$ could not be determined precisely.
There is the possibility that both the AFSC and SC states are found
for $x_{dSC}<x<0.09$, but~the SC solution will have lower energy. 
There is the AFSC-SC transition at $x=x_{dSC}$. 
The AFI state exists near half-filling for about $x<0.06$,
where doped holes form clusters and~localize.

The existence of AFI phase is closely related to the phase 
separation~\cite{arr91,kap12}
when the hole density is very small.  In~the phase-separated
phase, the~doped holes are localized and cannot be conductive.
The existence of AFI phase is determined by the quantity
\begin{equation}
\delta^2E(N_e)\equiv [E(N_e+\delta N_e)-2E(N_e)+E(N_e-\delta N_e)]/(\delta N_e)^2,
\end{equation}
where $E(N_e)$ is the ground-state energy with $N_e$ electrons.
$\delta^2 E(N_e)$ is approximately the second derivative of the
energy $E(N_e)$ and is proportional to the charge susceptibility.
When $\delta^2 E(N_e)$ is negative, the~phase separation occurs.
As shown in Figure~\ref{fig8}, the~phase separation occurs for $x<0.06$.
Concerning the phase separation, the~parameter $t'$ is important
because the phase separation region decreases as $-t'$ increases.
Thus,    the~AFI phase will decrease as $-t'$ increases.
The phase separation disappears for $t'=-0.2t$.

\section{Summary}\label{sec10}

We have discussed the possibility of high temperature superconductivity
in many-electron systems.  The~critical temperature $T_c$ may increase as
the characteristic energy of the interaction increases.
Empirically, $T_c$ is proportional to the inverse of the effective
mass of electrons.  $T_c$ is low when the effective mass is very heavy.
A candidate of high (room) temperature superconductivity may be
in materials with strong electron correlation and with small
effective mass enhancement. 
From this view point, the~repulsive Coulomb interaction can be a 
candidate of the origin of high temperature~superconductivity.

We have shown phase diagrams for the 2D Hubbard model and the
three-band d-p model.
The diagram in {Figure \ref{fig8}} exhibits the characteristic property of 
cuprate superconductors.
This supports that the origin of high temperature superconductivity
is the strong correlation between electrons.
That is, the~mechanism of high-$T_c$ superconductivity is the electron-pair
formation due to the strong on-site repulsive Coulomb interaction.
The competition between antiferromagnetism and superconductivity 
is important in realizing high temperature superconductivity.
High-$T_c$ superconductivity is expected in the region
near the boundary between AF phase and paramagnetic phase. 
In the phase diagram for the Hubbard model, the~SC phase exists
near the AF phase, and~AF order and superconductivity coexist
where the doping rate is approximately 0.05$\sim$0.06 $< x < x_{dSC}$
and  $0.08<x_{dSC}<0.09$.
We expect that this coexistence may be related to anomalous
metallic behavior in the underdoped region.  
The AF phase near half-filling is insulating, which is  approximately
for $x<0.06$.  There is the pure $d$-wave phase for $x>x_{dSC}$.

In the d-p model, the~AF region exists in the multi-dimensional
parameter space.  The~AF-PM boundary is a multi-dimensional 
region in this space.  Since we expect that
superconductivity occurs near the boundary, high temperature
superconductivity is more likely to occur in the d-p model.
There is the AF--PM boundary when the level difference $\Delta_{dp}$
is small. Thus, $T_c$ of high temperature cuprates will be high
when $\Delta_{dp}$ is small.  This tendency is consistent with
experimental $T_c$ of~cuprates. 

We give a comment on the crossover between weakly correlated
region and strongly correlated region.
We expect that this crossover is universal in the sense that
similar phenomena occur in nature.
There may be a universal class.  It will include the Kondo
effect~\cite{kon12,yan12,yan15}, QCD~\cite{ell96}, BCS-BEC crossover~\cite{noz85},
sine-Gordon model~\cite{raj89,sol79,yan16b,yan19b}, and~
Gross--Neveu model~\cite{gro74}.
\\
\\

\textbf{Acknowledgments}
This work was~supported by a Grant-in-Aid for Scientific
Research from the Ministry of Education, Culture, Sports, Science and
Technology of Japan (Grant No. 17K05559).
A part of the computations was supported by the Supercomputer
Center of the Institute for Solid State Physics, the~University of~Tokyo.

\end{document}